\pdfoutput=1
\RequirePackage{fix-cm}
\documentclass[smallextended]{svjour3}       
\smartqed  
\usepackage{graphicx}
\journalname{submitted}

\newcommand{\rmi}{\mathrm{i}}

\newcommand{\PC}{Physica C }

\newcommand{\JSNM}{J. Supercond. Nov. Magn. }
\newcommand{\PRB}{Phys. Rev. B }
\newcommand{\PRL}{Phys. Rev. Lett. }
\newcommand{\APL}{Appl. Phys. Lett. }
\newcommand{\JAP}{J. Appl. Phys. }
\newcommand{\SUST}{Supercond. Sci. Technol. }
\newcommand{\NatMat}{Nat. Mater. }

\begin{document}
\title{Directional vortex pinning at microwave frequency in YBa$_{2}$Cu$_3$O$_{7-x}$ thin films with BaZrO$_3$ nanorods.}

\titlerunning{Directional vortex pinning in YBa$_{-2}$Cu$_3$O$_{7-x}$ / BaZrO$_3$.}        

\author{N. Pompeo 	\and 
        K. Torokhtii	\and
        A. Augieri		\and 
        G. Celentano	\and
        V. Galluzzi	\and
        E. Silva
}

\institute{N. Pompeo   (\email{pompeo@fis.uniroma3.it}) 
            \and 
            K. Torokhtii 
            \and 
            E. Silva  (\email{silva@fis.uniroma3.it})
            \at
            Dipartimento di Fisica ``E. Amaldi'' and Unit\`a CNISM, Universit\`a Roma Tre, Via della Vasca Navale 84, 00146 Roma, Italy  
	    \and 
	    A. Augieri 
	    \and 
	    G. Celentano 
	    \and 
	    V. Galluzzi
	    \at
	    ENEA-Frascati, Via Enrico Fermi 45, 00044 Frascati, Roma, Italy
}

\date{May 10, 2012}

\maketitle

\begin{abstract}

We investigate the effect of the anisotropy and of the directional pinning in YBa$_2$Cu$_3$O$_{7-x}$ films grown by pulsed laser ablation from targets containing BaZrO$_3$ at 5\% mol. BaZrO$_3$ inclusions self-assemble to give nanorods oriented along the $c$-axis, thus giving a preferential direction for vortex pinning. The directionality of vortex response is studied at high ac frequency with the complex microwave response at 48 GHz, as a function of the applied field and of the angle $\theta$ between the field and the $c$-axis. The complex microwave response does not exhibit any angular scaling, suggesting that the structural anisotropy of YBa$_2$Cu$_3$O$_{7-x}$ is supplemented by at least another preferred orientation. The pinning parameter $r$ shows evidence of directional pinning, effective in a wide range of angles around the $c$-axis (thus ascribed to BZO nanocolumns).

\keywords{Vortex pinning \and Anisotropy \and Nanorods \and BaZrO$_3$}
\end{abstract}

\section{Introduction}
\label{sec:intro}

Vortex pinning in anisotropic superconductors is a long-standing issue. In fact, the inherent anisotropy of the compound determines by itself an anisotropic response, independently on pinning properties. In addition, coherence-length size structures give rise to vortex pinning. In YBa$_2$Cu$_3$O$_{7-x}$ (YBCO) it is well known \cite{kwok} that the layered crystal structure originates the so-called intrinsic pinning. In addition, extrinsic elongated structures such as BaZrO$_3$ (BZO) nanorods, which self-assemble in YBCO/BZO films grown by Laser Ablation, give another source of directional pinning. The interplay between directional pinning and structural anisotropy is still a matter of debate, and it is of particular interest \cite{maiorov,obradors,mcmanus} in view of the great potential of YBCO/BZO films and tapes for power applications (e.g., in coated conductors). In fact, BaZrO$_3$ (BZO) has been extensively studied \cite{maiorov,obradors,mcmanus,paturi,galluzziIEEE07,ENEA} due to the very strong pinning that it determines, either when it assembles in the shape of nanoparticles (typically in films grown by chemical methods) and when it forms nanorods instead (as in most films grown by Pulsed Laser Ablation). The typical transverse size of BZO nanorods is $\sim 5$ nm in diameter and $30-150$ nm in length, with the orientation approximately perpendicular to the film plane, that is along the $c$-axis \cite{galluzziIEEE07}. Thus, BZO nanorods are ideal extended defects for vortex pinning.

In ideally pinning-free anisotropic superconductors the mass tensor dictates the anisotropic behavior of the phyiscal properties. As an example, in cuprates a reduction of the dissipation takes place as the magnetic field approaches the $a,b$ planes due to the increase of the upper critical field $H_{c2}$, and the consequent smaller reduced field $H/H_{c2}$. 
More generally, the role of the mass anisotropy can be included with the so-called scaling property. Let $H_{c2}(\theta)=H_{c2}(0^{\circ})/\varepsilon(\theta)$ be the angle-dependent upper critical field in a uniaxial anisotropic superconductor, where $\theta$ is the angle between the field and the $c$-axis. The anisotropic Ginzburg-Landau treatment gives $\varepsilon^2(\theta)=\gamma^{-2}\sin^2\theta+\cos^2\theta$, 
where $\gamma=H_{c2}(90^{\circ})/H_{c2}(0^{\circ})$ is the anisotropy ratio ($\gamma\simeq 5\div 8$ in YBCO).
It is theoretically known \cite{blatterPRL,haoclem} that in the London approximation the thermodynamic and intrinsic transport properties obey the following scaling behavior: the field and angle dependencies are only through the reduced field $H/H_{c2}(\theta)$. Thus, measurements taken at different fields and angles and plotted as a function of $H\varepsilon(\theta)$ are expected to collapse on a single curve. It is an important point that such scaling is grounded on the existence of only one field scale: extrinsic properties, such as pinning, may depend on different field scales (e.g., the matching field in samples with regular lattices of defects), and extrinsic preferential directions can exist (e.g., columnar defects at a specific angle), thus breaking the scaling property. It is worth noting that point pinning, on the other hand, preserves the scaling. In the context of cuprate materials for applications it has proven useful to introduce the above-mentioned scaling with some effective anisotropic ratio \cite{obradors}, in order to take into account the reduced anisotropy in the transport properties in samples with a high density of artificially introduced defects.

In this paper we present measurements of the response of the vortex matter to a microwave field at 48 GHz in YBCO thin films with BZO nanorods. This approach brings important information since only very short-range vortex oscillations are induced, thus avoiding the complications  originating from the long-range vortex dynamics as probed by dc resistivity and critical current. We will show that even in this regime, the angular scaling never applies. Instead, we present evidence that correlated pinning exists and dominates the angular response even at our very high driving frequency. 

The paper is organized as follows: in Section \ref{sec:technique} we describe the experimental technique and the sample studied. In Section \ref{sec:ang} we present the experimental demonstration of the absence of angular scaling. In Section \ref{sec:dirpin} we present and shortly discuss the evidence for correlated pinning.

\section{Measurement technique and samples}
\label{sec:technique}
We have measured the complex microwave response  by means of a dielectric loaded cylindrical cavity excited in the TE$_{011}$ mode, with a resonant frequency $f_0\simeq$47.7 GHz at cryogenic temperatures. The sample replaces one of the cavity end-wall surfaces, where the TE$_{011}$ mode induces a circular pattern of microwave currents, and therefore contributes to the resonator unloaded quality factor $Q$ and resonant frequency $f_0$. Only planar currents are induced in the sample. The resonator has been described in detail elsewhere \cite{pompeoJS07}.
The field and angle shift of $Q$ and $f_0$ are related to the complex vortex resistivity $\Delta\tilde\rho_v(H,\theta)=\tilde\rho_v(H,\theta)-\tilde\rho_v(0,0)$ by the well-known relations:
\begin{eqnarray}
\label{eq:rhoeff}
\delta \tilde{z}=\left(\frac{1}{Q(H,\theta)}-\frac{1}{Q(0,0)}\right)-2\mathrm{i}\frac{f_0(H,\theta)-f_0(0,0)}{f_0(0,0)}=\frac{\Delta\tilde\rho_v(H,\theta)}{Gd}
\end{eqnarray}
\noindent where $\delta \tilde{z}$ is the adimensional complex microwave response, $G$ is a geometrical factor and $d$ is the film thickness. In building up Eq.(\ref{eq:rhoeff}) we made use of the so-called thin-film approximation \cite{stratiSTOnoi}, well verified with the thickness of our samples. In the present paper we are not interested in the absolute value of the complex resistivity, and the data will be presented as $\Delta(1/Q)$ and $\Delta f_0/f_0$.

YBCO $c$-axis epitaxial films have been grown by Pulsed Laser Ablation starting from targets containing BZO powders at various \% mol. on SrTiO$_3$ substrates \cite{galluzziIEEE07}. The films are squares with 7.5 mm side and $d$=120 nm thickness, which places them within the above mentioned thin film approximation. A complete dc and microwave characterization has been reported elsewhere \cite{galluzziIEEE07,pompeoAPL07,pompeoJAP09}. We here focus on the results obtained on one YBCO/BZO sample at 5\% mol.
All the measurements were performed at selected temperatures, carefully chosen in order to avoid the substrate resonances caused by SrTiO$_3$ \cite{stratiSTOnoi} .
The BZO inclusions generated columnar-like defects, approximately perpendicular to the film plane, as observed with transverse TEM images \cite{augieriJAP10}, while such defects were not detected in pure YBCO samples. The measured $T_c$ (zero dc resistance criterion) was consistently $\sim$90 K in all samples.

\section{Failure of the angular scaling}
\label{sec:ang}
In this Section we present the data of the microwave response taken at different angles, and we show that no angular scaling takes place, even if an effective anisotropy function is introduced.
\begin{figure}[h]
\centerline{\includegraphics[width=12cm]{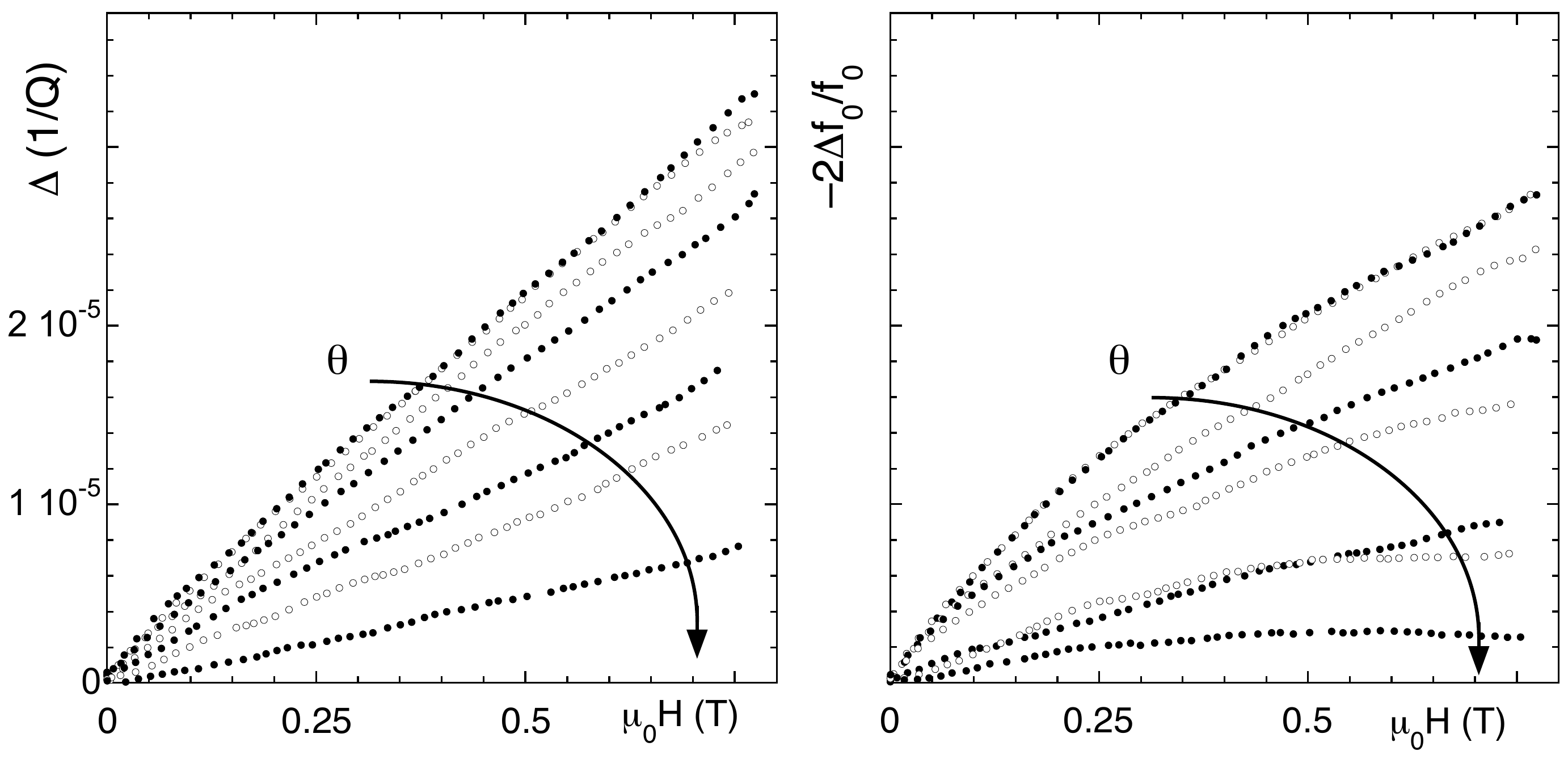}}
  \caption{Field sweeps of $\Delta Q^{-1}(H,\theta)$ and $(-2)\Delta f_0(H,\theta)/f_0$, left and right panel respectively; $T$=81 K, selected angles ($\theta$=0, 10, 30, 45, 60, 75, 80, 90 degrees).}
\label{fig:raw}
\end{figure}

The data have been collected at the fixed temperature $T=81\;$K. Field sweeps were performed for different field orientations. Fig. \ref{fig:raw} reports the data for the adimensional complex microwave response $\delta\tilde{z}$ at selected angles $\theta$. It can be seen that, by increasing the tilting angle from the normal orientation ($H//c$-axis) to the parallel orientation ($H$//$ab$-plane), $\delta\tilde{z}$ progressively decreases both in the real and in the imaginary parts. 

\begin{figure}[h]
\centerline{\includegraphics[width=12cm]{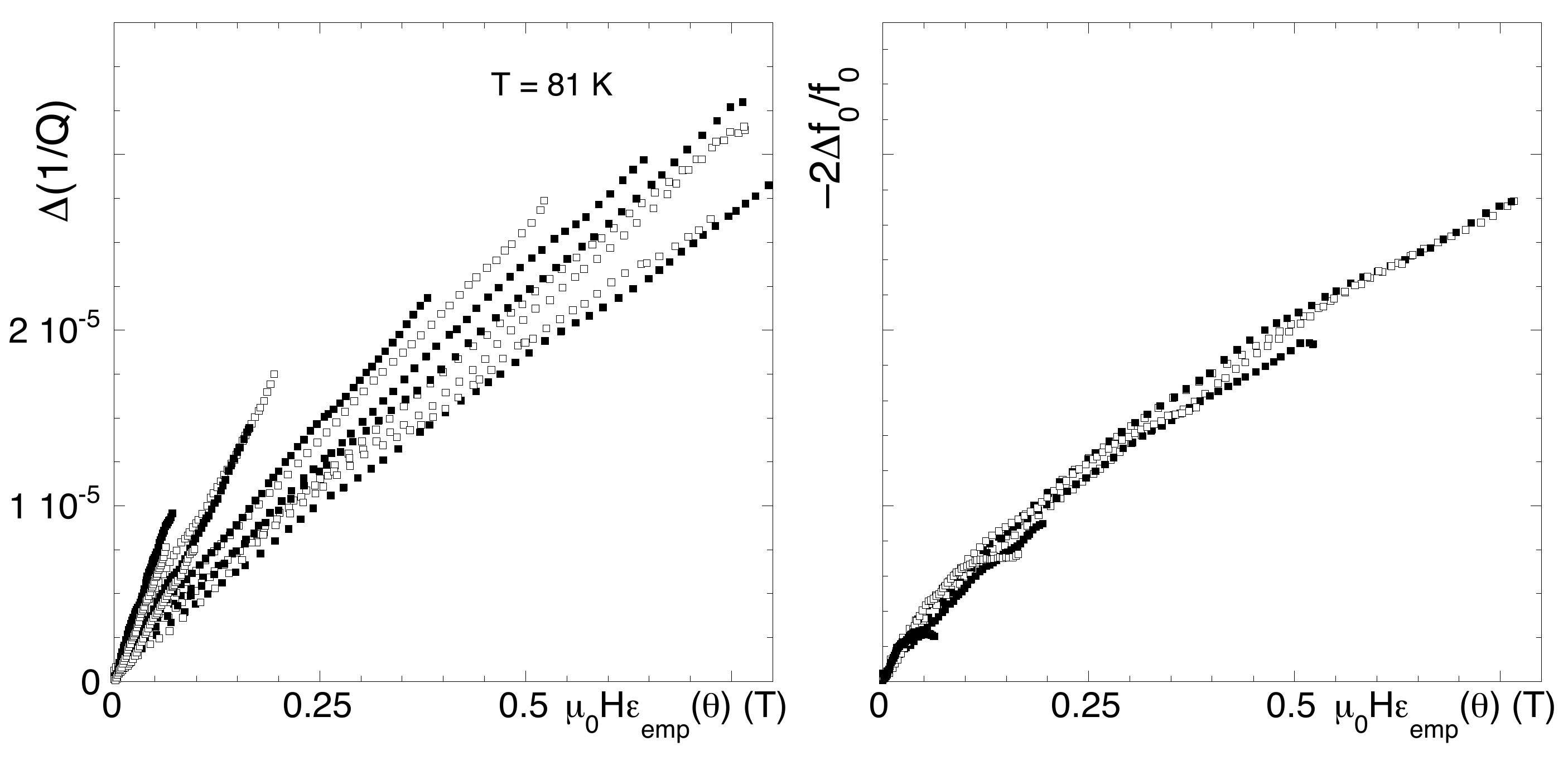}}
  \caption{Same data of Fig.\ref{fig:raw}, plotted versus a rescaled field with an empirically  scaling function $\varepsilon_{emp}(\theta)$, determined in order to achieve best scaling. As it can be seen, it is not possible to scale both the real and imaginary part of the adimensional complex microwave response: here we demonstrate that, by scaling the imaginary part (right panel), the real part (left panel) does not scale.}
\label{fig:noscaling}
\end{figure}

To test whether an angular scaling takes place or not, we have tried to scale the curves with some empirical scaling function $\varepsilon_{emp}(\theta)$, that is we rescaled the applied field of each curve in order to make it overlap to the curve taken at $\theta=0$. We have chosen to use an empirical function $\varepsilon_{emp}(\theta)$ in order to accept, for instance, some ``effective" anisotropy, and more generally to release some possibly too tight theoretical constraint. However, even in this somewhat less rigorous version, we could not find a proper scaling of our data: in Fig.\ref{fig:noscaling} we report an example of the result of such a procedure. While it is possible to scale the imaginary or the real part of $\delta\tilde{z}$, it is never possible to scale both $\Delta(1/Q)$ and $\Delta f/f_0$. 

Nevertheless, in order to definitely assess the issue, we have to notice a possible source of complications in our experimental configuration. In fact, due to the circular path of the microwave currents, by varying the angle $\theta$ one varies also the Lorentz force acting on flux lines. This effect can in principle affect the angular scaling (even if in very anisotropic superconductors such as Bi$_2$Sr$_2$CaCu$_2$O$_{8+x}$ the angular scaling takes place also with varying Lorentz force \cite{silvaPRB97}), and consequently it  requires careful consideration. Indeed, by tilting the field away from the perpendicular orientation, fluxons form angles $\neq90^\circ$ with the currents, depending on the position on the sample surface. As a consequence, the overall effect has to be averaged over the sample surface. On very general grounds, the resistivity incorporates a factor proportional to the electric field originating from the Lorentz force $\mathbf{J}\times \mathbf{B}$ \cite{Tinkhambook,GR}. This factor can acquire a very complicated angular dependence in anisotropic superconductors \cite{HHT,noifuturo}, but the essential point is that this additional factor is just an additional  mutiplying function for the resistivity. Thus, the vortex complex resistivity as a function of the field and angle $\theta$ in a varying Lorentz force configuration, can be cast in the form:
\begin{eqnarray}
\label{eq:rholor}
\Delta\tilde\rho_{v}(H,\theta)=f_L(\theta)\Delta\tilde\rho_{v,MLF}(H,\theta)
\end{eqnarray}
where  $f_L(\theta)=1$ if $\mathbf{B}\perp\mathbf{J}$. Here, $\Delta\tilde\rho_{v,MLF}(H,\theta)$ represents the angle-dependent vortex resistivity with constant (maximum) Lorentz force. The latter, should the angular scaling take place, would scale as $\Delta\tilde\rho_{v,MLF}(H\varepsilon_{emp}(\theta))$, where again we have included the possibility of an ``effective" anisotropy function, as before.

\begin{figure}[h]
\centerline{\includegraphics[width=6cm]{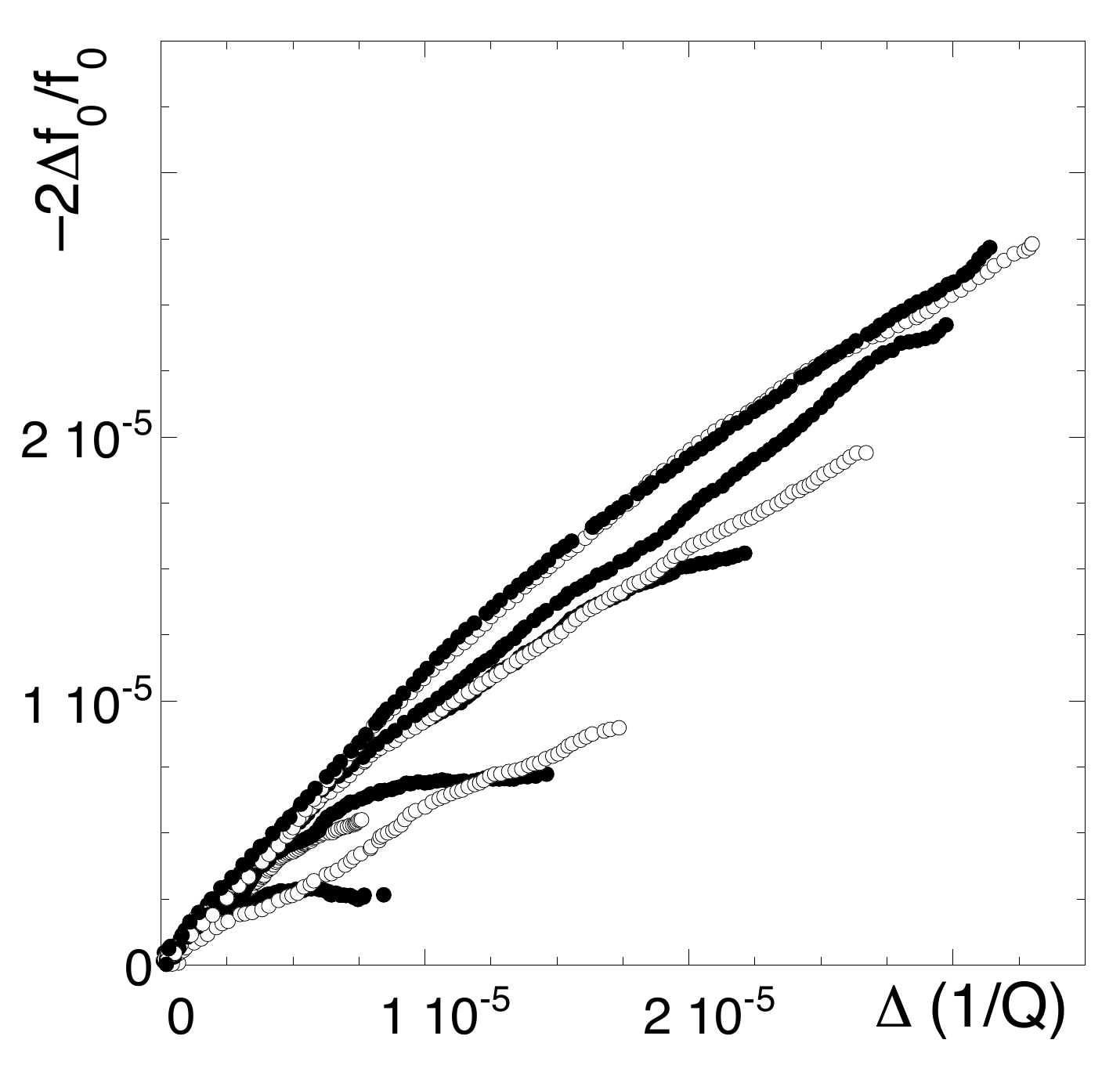}}
  \caption{Parametric plot of the raw data of Fig.\ref{fig:raw}, $(-2)\Delta f_0(H,\theta)/f_0$ vs $\Delta Q^{-1}(H,\theta)$, with $H$ as running parameter and selected angles ($\theta$=0, 10, 30, 45, 60, 75, 80, 90 degrees); $T$=81 K.}
\label{fig:xvsr}
\end{figure}

In order to exhaustively test the possible scaling of $\Delta\tilde\rho_{v}$ (which, we recall, would require the simultaneous scaling of both its real and imaginary parts) without the need of determining  $f_L(\theta)$, we consider a plot of $-2\Delta f_0/f_0$ versus $\Delta(1/Q)$, which is a representation largely used in literature \cite{maeda}. In Fig. \ref{fig:xvsr} several curves at different $\theta$ (raw data from Fig.\ref{fig:raw}) are plotted with $H$ as a running parameter. 
It is clearly seen that the various curves do not collapse together. In particular, their steepness depend on the angle $\theta$. This is the final demonstration that the vortex complex resistivity does not exhibit angular scaling in YBCO/BZO with nanorods: in fact, even if the Lorentz-force contribution $f_L(\theta)$ is not explicitly known, when $-2\Delta f_0/f_0 \propto \Delta(1/Q)$, as in the present case in a large part of the curves, the parametric plot of Fig.\ref{fig:xvsr} is insensitive to $f_L(\theta)$, which is merely a multiplying factor on both axis, and thus does not change the steepness of $(-2)\Delta f_0(H,\theta)/f_0$ vs $\Delta Q^{-1}(H,\theta)$.\footnote{We note that the same can be said of the geometrical factors $G$ and $d$, see Eq.(\ref{eq:rhoeff}), which makes our conclusions fully supported by the simple analysis of the raw data for the complex resistivity shift.}

Thus, angular scaling does not take place in the complex resistivity in YBCO/BZO, even taking into account a possible effective anisotropy function. This is the first experimental result of this paper. It is quite natural to ascribe the failure of the scaling to the directional pinning due to nanorods, issue that we address in the next section.

\section{Evidence for directional pinning.}
\label{sec:dirpin}

In order to keep as much contact as possible to the raw data, we discuss vortex pinning at our microwave frequency by using the well-known adimensional vortex pinning parameter $r$, which is defined directly from the raw data for the adimensional complex microwave response as:
\begin{equation}
\label{eq:rraw}
    r=
    \left[{-2\frac{\Delta f_0(H,\theta)}{f_0}}\right]
    /
    \left[
    \Delta\frac{1}{Q(H,\theta)}
    \right]
\end{equation}
\noindent which does not depend on any geometrical factor relating the adimensional complex microwave response and the complex vortex resistivity (see Eq.(\ref{eq:rhoeff})). The physical meaning is obvious from Eq.(\ref{eq:rhoeff}): $r$ represents the balance between the out-of-phase (elastic) and in-phase (dissipative) responses of the vortices to the alternating driving field. The larger the pinning, the larger $r$. One may make a step further, introducing e.g. the established vortex motion model by Gittleman and Rosenblum \cite{GR} (to which, by the way, many different models correspond when vortex creep can be neglected \cite{pompeoPRB08}), where the complex resistivity can be written:
\begin{equation}
\label{eq:rhoGR}
    \Delta\tilde\rho_{v,GR}=\rho_{ff}\frac{1}{1-\rmi\omega_p/\omega}
\end{equation}
\noindent where $\rho_{ff}$ is the flux flow resistivity and it includes the Lorentz force correction given by Eq.(\ref{eq:rholor}), $\omega=2\pi f$ is the microwave angular frequency and $\omega_p$ is the (de)pinning frequency. For $\omega>\omega_p$ the complex resistivity is insensitive to pinning. Within this model, $r=\omega_p/\omega$. We stress that within this model $r$ is also independent on the Lorentz force contribution $f_L(\theta)$ introduced in the previous Section.
\begin{figure}[h]
\centerline{\includegraphics[width=6cm]{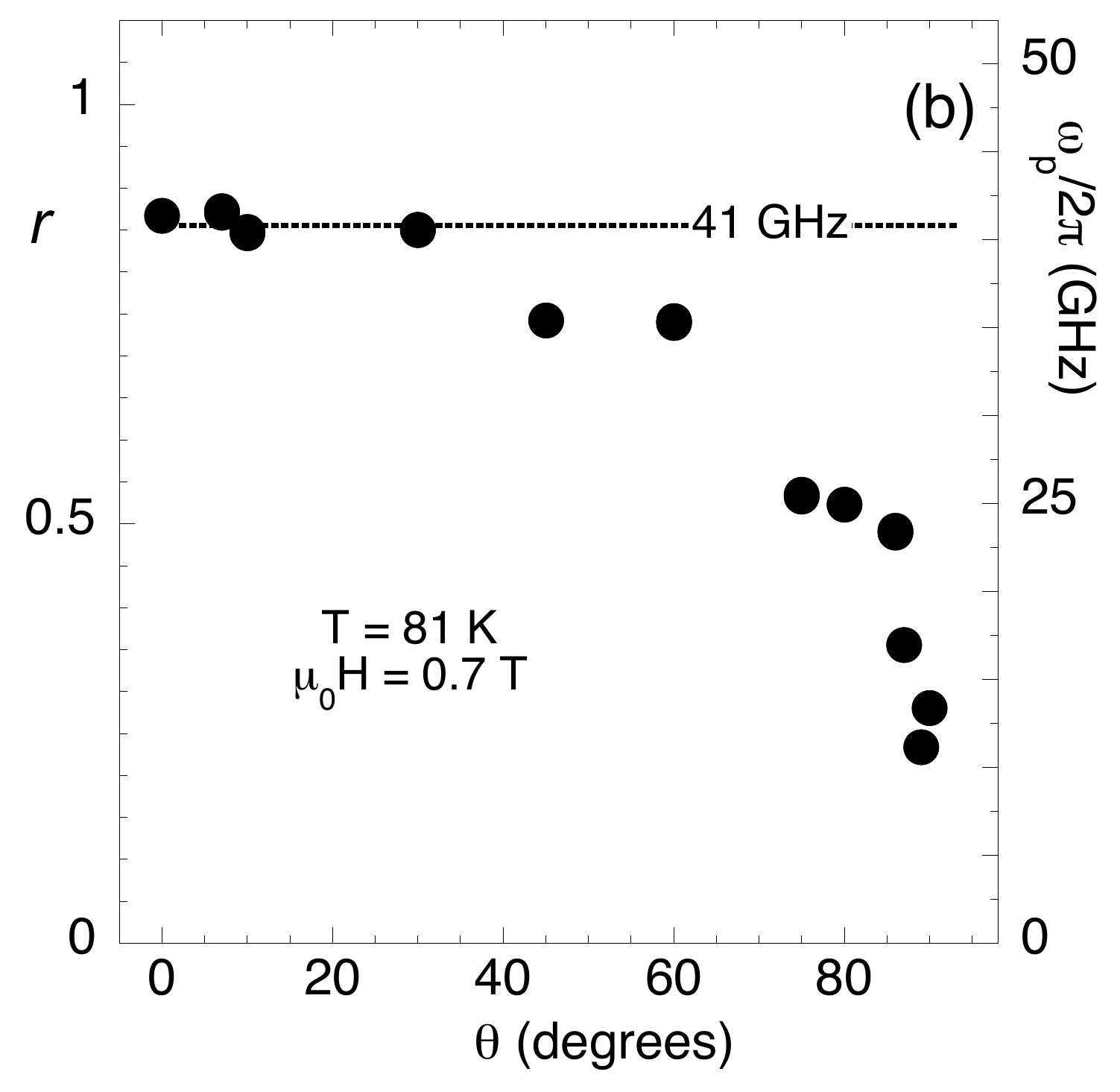}}
  \caption{Pinning parameter $r$ vs. $\theta$ at $\mu_0H=$0.7 T. The right scale reports the corresponding depinning frequency $\omega_p/2\pi$. It is clearly seen that pinning is much stronger when the field is close to the c axis. $T$=81 K.}
\label{fig:r}
\end{figure}

The microwave investigation of anisotropic pinning in pure YBCO has received some attention in the past. It has been observed that, apart from a very narrow (and somewhat elusive) peak in $\omega_p$ when $H // (a,b)$ (here, $\theta=90^\circ$), at a fixed field $\omega_p$ is nearly constant with $\theta$ \cite{golos}. This behavior is consistent with the absence of any directional pinning, apart the very narrow angular range at $\theta=90^\circ$ where intrinsic pinning has some influence. Moreover, values of $\omega_p/2\pi\simeq$10 GHz are commonly reported \cite{golos}. In order to observe the angular dependence of the vortex pinning, we calculated the pinning parameter from our data near to our maximum field, $\mu_oH=$0.7 T. In Fig.\ref{fig:r} we report the so-obtained $r$ vs. $\theta$. We do not have evidence of the intrinsic pinning peak at $\theta=90^\circ$: it can be due to the limited angular resolution of the measurements here presented, or to the thermal weakening of the intrinsic pinning at our rather high $T=$81 K.  The behavior in parallel orientation will be left to a future investigation. We focus here on two important features for our purposes: first, $r$ attains very large values, $r\sim1$, thus indicating very strong pinning. Second, $r$ decreases with increasing $\theta$, demonstrating that pinning is much stronger when $H // c$. In order to quantify the pinning effect, in the same figure the left vertical scale reports the depinning frequency obtained assuming $r=\omega_p/\omega$. It is seen that very large $\omega_p/2\pi\simeq$41 GHz is obtained in a wide angular range $\theta\leq30^\circ$. We note that such high $\omega_p$ is  larger than the value reported in YBCO with columnar defects introduced by heavy-ion irradiation \cite{SilvaJLTP2003}. When the angle is tilted further, $\omega_p$ decreases and attains values $\sim$ 15 GHz, in agreement with common reports \cite{golos}.

The results here presented can be well understood when the morphology of our YBCO/BZO sample is taken into account: in fact, TEM images clearly show the existence of correlated defects approximately parallel to the c-axis \cite{augieriJAP10}. Thus, the behavior of vortex pinning as measured by $r$ is fully consistent with very large correlated pinning along the $c$ axis. This result, together with the failure of the angular scaling, points to an angular dependence of the raw data for the electrical transport properties that is largely determined by BZO-induced correlated defects.

\section{Summary}
\label{conc}
We have presented angular measurements of the vortex microwave complex response in a YBCO film with BZO nanorods. We have experimentally demonstrated that no angular scaling takes place, even allowing for a scaling function different from the theoretical prescription. We have shown that the pinning parameter exhibits a large increase when the field is directed along the nanorods preferential direction. The results demonstrate that the angular dependence of the raw data for the electrical transport properties is dictated by the orientation of the nanorods.
\begin{acknowledgements}
We thank S. Schweizer for the help in taking data. This work has been partially supported by the FIRB project ``SURE:ARTYST" and by EURATOM. N.P. acknowledges support from Regione Lazio.
\end{acknowledgements}


\begin{thebibliography}{}
%
\bibitem{kwok} W. K. Kwok, U. Welp, V. M. Vinokur, S. Fleshler, J. Downey, and G. W. Crabtree, \PRL {\bf 67}, 390 (1991)
%
\bibitem{maiorov} B. Maiorov, S. A. Baily, H. Zhou, O. Ugurlu, J. A. Kennison, P. C. Dowden, T. G. Holesinger, S. R. Foltyn, and L. Civale, \NatMat {\bf 8} 398 (2009).
%
\bibitem{obradors} J. Guti\'errez, A. Llord\'es, J. G\'azquez, M. Gibert, N. Rom\`a, S. Ricart, A. Pomar, F. Sandiumenge, N. Mestres, T. Puig and X. Obradors, \NatMat {\bf 6}, 367 (2007).
%
\bibitem{mcmanus} J. L. Macmanus-Driscoll, S. R. Foltyn, Q. X. Jia, H. Wang, A. Serquis, L. Civale, B. Malorov, M. E. Hawley, M. P. Maley, and D. E. Peterson, \NatMat {\bf 3}, 439 (2004).
%
\bibitem{paturi} M. Peurla, H. Huhtinen, M. A. Shakhov, K. Traito, Yu P. Stepanov, M. Safonchik, P. Paturi, Y. Y. Tse, R. Palai, and R. Laiho, \PRB {\bf 75}, 184524 (2007).
%
\bibitem{galluzziIEEE07} V. Galluzzi, A. Augieri, L. Ciontea, G. Celentano, F. Fabbri, U. Gambardella, A. Mancini, T. Petrisor, N. Pompeo, A. Rufoloni, E. Silva, and A. Vannozzi, IEEE Trans. Appl. Supercond. {\bf 17}, 3628 (2007)
%
\bibitem{ENEA} A. Augieri, V. Galluzzi, G. Celentano, A. A. Angrisani, A. Mancini, A. Rufoloni, A. Vannozzi, E. Silva, N. Pompeo, T. Petrisor, L. Ciontea, U. Gambardella, and S. Rubanov, IEEE Trans. Appl. Supercond. {\bf 19}, 3399 (2009)
%
\bibitem{blatterPRL} G. Blatter, V B. Geshkenbein, and A. I. Larkin, \PRL {\bf 68}, 876 (1992).
%
\bibitem{haoclem} Z. Hao and J. R. Clem, \PRB {\bf 46}, 5853 (1992).
%
\bibitem{pompeoJS07} N. Pompeo, R. Marcon, and E. Silva, \JSNM {\bf 20}, 71 (2007).
%
\bibitem{stratiSTOnoi} N. Pompeo, L. Muzzi, V. Galluzzi, R. Marcon, and E. Silva, \SUST {\bf 20}, 1002 (2007).
%
\bibitem{pompeoAPL07} N. Pompeo, R. Rogai, E. Silva, A. Augieri, V. Galluzzi, and G. Celentano, \APL {\bf 91}, (2007) 182507.
%
\bibitem{pompeoJAP09} N. Pompeo, R. Rogai, A. Augieri, V. Galluzzi, G. Celentano, and E. Silva, \JAP {\bf 105}, 013927 (2009).
%
\bibitem{augieriJAP10} A. Augieri, G. Celentano, V. Galluzzi, A. Mancini, A. Rufoloni, A. Vannozzi, A. Angrisani Armenio, T. Petrisor, L. Ciontea, S. Rubanov, E. Silva, and N. Pompeo, \JAP {\bf 108}, 063906 (2010).
%
\bibitem{silvaPRB97} E. Silva, S. Sarti, M. Giura, R. Fastampa and R. Marcon, \PRB {\bf 55}, 11115 (1997).
%
\bibitem{Tinkhambook} M. Tinkham, {\it Introduction to Superconductivity, 2nd Ed.}, Dover Publications, 2004.
%
\bibitem{GR} J. Gittleman and B. Rosenblum, \PRL {\bf 16}, 734 (1966).
%
\bibitem{HHT} Z. Hao, C-Ren Hu, C.-S. Ting, \PRB  {\bf 51}, 9387 (1995); \PRB {\bf 52}, R13138 (1995).
%
\bibitem{noifuturo} N. Pompeo and E. Silva, in preparation.
%
\bibitem{maeda} A. Maeda, Y. Tsuchiya, K. Iwaya, K. Kinoshita, T. Hanaguri, H. Kitano, T, Nishizaki, K. Shibata, N. Kobayashi, J. Takeya, K. Nakamura and Y. Ando, \PC {\bf 362}, 127 (2001)
%
\bibitem{pompeoPRB08} N. Pompeo and E. Silva, Phys. Rev. B {\bf 78} (2008) 094503.
%
\bibitem{golos} G. Golosovsky, M. Tsindlekht and D. Davidov, \SUST {\bf 9}, 1 (1996), and references therein.
%
\bibitem{SilvaJLTP2003} E. Silva, R. Marcon, R. Fastampa, M. Giura, S. Sarti, G. Ghigo, J. Low Temp. Phys. {\bf 131}, 871 (2003).
%
\end{thebibliography}
\end{document}